\documentclass[twocolumn,floatfix]{aastex631}
\usepackage[utf8]{inputenc}
\usepackage{float}
\pdfoutput=1
\shorttitle{Abell 2261 in a transitional dynamical state}
\shortauthors{Kim et al.}

\graphicspath{{./}{figures/}}

\begin{document}

\title{Is Abell 2261 a fossil galaxy cluster in a transitional dynamical state?}

\author[0000-0003-4032-8572]{Hyowon Kim}
\affiliation{Korea Astronomy and Space science Institute, 776, Daedeokdae-ro, Yuseong-gu Deajeon, 34055, Korea}
\affiliation{Korea University of Science and Technology, 217, Gajeong-ro, Yuseong-gu Deajeon, 34113, Korea}

\author{Jongwan Ko}
\affiliation{Korea Astronomy and Space science Institute, 776, Daedeokdae-ro, Yuseong-gu Deajeon, 34055, Korea}
\affiliation{Korea University of Science and Technology, 217, Gajeong-ro, Yuseong-gu Deajeon, 34113, Korea}

\author{Rory Smith}
\affiliation{Korea Astronomy and Space science Institute, 776, Daedeokdae-ro, Yuseong-gu Deajeon, 34055, Korea}
\affiliation{Korea University of Science and Technology, 217, Gajeong-ro, Yuseong-gu Deajeon, 34113, Korea}

\author{Jae-Woo Kim}
\affiliation{Korea Astronomy and Space science Institute, 776, Daedeokdae-ro, Yuseong-gu Deajeon, 34055, Korea}

\author{Ho Seong Hwang}
\affiliation{Department of Physics and Astronomy, Seoul National University, 1 Gwanak-ro, Gwanak-gu, Seoul 08826, Korea}
\affiliation{SNU Astronomy Research Center, Seoul National University, 1 Gwanak-ro, Gwanak-gu, Seoul 08826, Korea}

\author{Hyunmi Song}
\affiliation{Department of Astronomy and Space Science, Chungnam National University, Daejeon 34134, Republic of Korea}

\author{Jihye Shin}
\affiliation{Korea Astronomy and Space science Institute, 776, Daedeokdae-ro, Yuseong-gu Deajeon, 34055, Korea}

\author{Jaewon Yoo}
\affiliation{Korea Astronomy and Space science Institute, 776, Daedeokdae-ro, Yuseong-gu Deajeon, 34055, Korea}
\affiliation{Korea University of Science and Technology, 217, Gajeong-ro, Yuseong-gu Deajeon, 34113, Korea}



\begin{abstract}

Abell 2261 (A2661) is a well-studied fossil cluster, but previous studies give contradictory results on its dynamical states, such as its X-ray central entropy and magnitude gap. To improve our understanding of its dynamical state, we conduct multi-object spectroscopic observations with Hectospec on the MMT, covering an area out to 5 virial radii from the cluster center, and get improved completeness and membership. Using this new data, we calculate multiple dynamical indicators including gaussianity, distance offset, and velocity offset. These indicators suggest that A2261 is moderately relaxed. However, Dressler-Shectman test reveals a group candidate to the south, at a projected distance that is near the virial radius, and that overlaps with an X-ray tail-like feature. One of the galaxies associated with that group would be sufficiently bright to reduce the fossil magnitude gap. This raises the possibility that A2261 could have recently transit in fossil status, if the group had previously crossed the cluster and is only now found outside. In the cluster outskirts, we see an extended feature of galaxies located on the opposite side of the cluster to the group candidate. On even larger scales, we find this feature connects, both on the sky and in velocity space, with a long ($\sim$4.4Mpc) filamentary structure in the Sloan Digital Sky Survey data. This could support the idea that a group was fed into the cluster through the filament, temporarily breaking the fossil status, and resulting in a minor merger that weakly disturbed the intracluster medium of the cluster.

\end{abstract}

\keywords{\href{http://astrothesaurus.org/uat/9}{Abell clusters (9)},  
\href{http://astrothesaurus.org/uat/584}{Galaxy clusters (584)},
\href{http://astrothesaurus.org/uat/591}{Galaxy Dynamics (591)},
\href{http://astrothesaurus.org/uat/600}{Galaxy Interactions (600)} }


\section{Introduction} \label{sec:intro}
Galaxy clusters are the most massive and the largest gravitationally bound systems in the Universe. According to the hierarchical structure formation scenario, naturally, many gravitational interactions inside of the cluster halo bring about the formation of the brightest cluster galaxy (BCG) at the gravitational center. Over time, massive galaxies are strongly affected by dynamical friction, and the BCG continues to merge with massive galaxies. Within this simple scenario, the observed properties of galaxy clusters should reflect the history of the evolution of the cluster.

In some clusters, it is predicted that a major component of the cluster's mass is formed in an early epoch of the universe and since then they passively evolve through mainly minor mergers \citep[e.g.][]{1994Natur.369..462P, 2005ApJ...630L.109D, 2010MNRAS.405.1873D}. If we accept this simple scenario, one obvious prediction is that older and more relaxed galaxy systems will present large differences in brightness between their BCG and their cluster member galaxies, and these are known as `fossil systems'. \citet{2003MNRAS.343..627J} defined this fossil galaxy system as having a large magnitude gap, \textit{$\Delta M_{12}\geq2$ (in r band)}, between the BCG and the second brightest cluster galaxy within the half virial radius, and with high X-ray luminosity, \textit{$L_{x} > 10^{42} erg s^{-1}$}. Fossil clusters are thought as objects at the high mass end of those oldest and undisturbed galaxy systems \citep[e.g.][]{1994Natur.369..462P,2003MNRAS.343..627J,2006RMxAC..26..111C}. 

The large magnitude gap feature of fossil systems has been studied extensively to determine their origins. The reason for the large magnitude gap had been thought as the special characteristics of a fossil system's BCG. It was expected that fossil BCGs would be more massive than other BCGs in non-fossil systems, compared to their cluster mass. However, many studies have found that the properties of fossil BCGs are, in fact, not significantly different from normal BCGs \citep[e.g.][]{2006MNRAS.369.1211K,2009AJ....137.3942L}. Other works were focused on their member galaxies. They found that there was a lower fraction of galaxies brighter than M* magnitude in fossil systems compared to in normal systems. Therefore, the main hypothesis is that M* galaxies were swallowed by BCG growth inside of galaxy groups \citep{2006AJ....131..158M, 2014A&A...565A.116Z, 2015A&A...581A..16Z}. 

Other studies have attempted to understand if the magnitude gap between a BCG and the second brightest galaxy is a good criterion enough to select early-formed and relaxed systems such as fossils \citep{2010MNRAS.405.1873D,2016ApJ...824..140R}. There is much debate about whether the concept of a fossil is a final stage of dynamical state or a more transitional event, by studying the magnitude gap difference \citep{2008MNRAS.386.2345V,2010MNRAS.409..169S,2010MNRAS.405.1873D,2015A&A...581A..16Z}. \citet{2008MNRAS.386.2345V} argued that whenever a galaxy system has a large magnitude gap, the difference can be updated by continuous matter infalling from the surrounding environment. \citet{2010MNRAS.405.1873D} selected fossil groups at three different redshifts using simulations, and traced them until z=0, but none of them maintained their large magnitude gap after $\sim$4Gyr. \citet{2015A&A...581A..16Z} shows the existence of substructures in a cluster-mass fossil system. They suggest that fossil clusters can have different origins, and the fossil condition at cluster masses is not firmly relaxed. 

We select Abell 2261 (A2261) for study as a system with a contradictory dynamical state, and may be an example of the disturbed fossil system whose origin we hope to better understand. In \citet{2017ApJ...836..105K}, a good correlation between magnitude gap and X-ray central entropy is found (see their Figure 2). However, in their figure, there are two outlier points, which are scattered from the main tendency. These outlier clusters have a controversial dynamical state that high magnitude gap but with a substantial X-ray central entropy value \citep{2009ApJS..182...12C}. Although both are outliers, A2261 was chosen because it fulfills the conventional fossil criteria, thus it can show a more prominent hint than the other. Using wider and more complete multi-object spectroscopic observations, we revisit A2261 to better understand its controversial dynamical state. 

In Section \ref{sec:data}, we describe our spectroscopic observations and data reduction, catalog data, and cluster member selection. We then show results about the dynamical state of A2261 in Section \ref{sec:rean}. Using new data, we check simple dynamical indicators, Dressler$-$Shectman test \citep[D-S test;][]{1988AJ.....95..985D}, and compare spectroscopic membership position with X-ray data. We also test the membership probability of galaxies in the cluster by comparison with simulation data, and look at filament connections with the cluster using the Sloan Digital Sky Survey \citep[SDSS;][]{2000AJ....120.1579Y} data. In Section \ref{sec:dis}, we discuss the possible scenarios for the dynamical state of A2261 and the possibility of a transitional origin for some fossil clusters. Section \ref{sec:sumncon} presents a summary of our study.
In this paper, we adopt a $\Lambda$CDM cosmology with $H_{0}=70km s^{-1} Mpc^{-1}$, $\Omega_{m}=0.3$ and $\Omega_{\Lambda}=0.7$, which give a plate scale of 3.604kpc/$\arcsec$ at the redshift (z=0.2242) for A2261.

\section{The data} \label{sec:data}
\begin{figure*}[htp]
\begin{center}
\includegraphics[scale=.40]{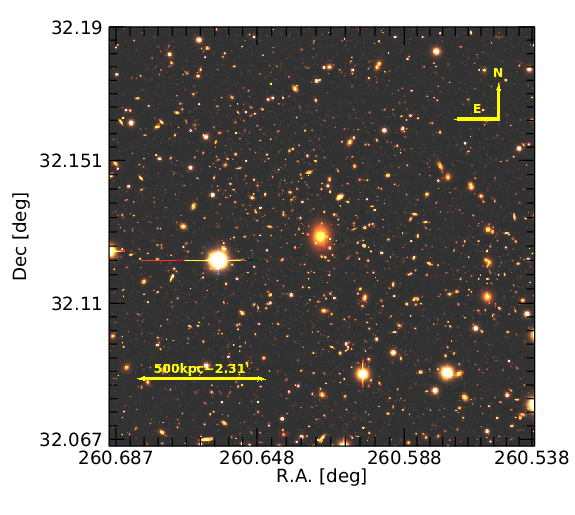}
\caption{Subaru Suprime-Cam color image for the central 5.6$\arcmin\times$5.7$\arcmin$ ($\sim$1Mpc$\times$1Mpc) region of A2261. It shows the BCG dominant features of the cluster. 500Kpc is $\sim$ 2.31 arcmin at z=0.2242.}\label{fig:fig1}
\end{center}
\end{figure*}

\begin{table*}[htp]
 \caption{Physical properties of A2261.}\label{tab:a2261}
  \begin{tabular}{cccccccc} 
  \hline
    Name & R.A. & Dec & Redshift &  M$_{vir}$ & ${R_{200}}^{\alpha}$ & ${\sigma_{cl}}^{\alpha}$ & ${L_{x}}^{\beta} $ \\[0.5ex]
      & (J2000) & (J2000) & & [Mpc] & $[10^{14} M_{\sun}]$& $[10^{3}km s^{-1}]$&$[ergs^{-1}]$\\[0.5ex]
    \hline    
      A2261 & 260.6129 & 32.1338 & 0.2242 &  2.03 $\pm$ 0.23 & 1.114 & 0.663 & 5.55 x$10^{43}$ \\[0.5ex]
   \hline
   \end{tabular}
 \tablecomments{ R.A., Dec, Redshift and M$_{vir}$ are from \citet{2013ApJ...767...15R}. $^{\alpha}$ ${R_{200}}$ and $ {\sigma_{cl}}$ is based on our own measurements. $^{\beta}$ ${L_{x}}$ is from the ACCEPT project \citep{2009ApJS..182...12C}. }
\end{table*}

\subsection{Photometric data} \label{subsec:survey}
To choose targets for multi-object spectroscopy, the SDSS Data Release 12 \citep[DR12;][]{2015ApJS..219...12A} was used. The composite model magnitude (\texttt{cModelMag}) is derived from a linear combination of the exponential and the de Vaucouleurs fitting method, thus it is better than others to fit the galaxy light profile. The model magnitude (\texttt{modelMag}) is obtained by fitting with a higher likelihood model to measure the flux through equivalent apertures for the unbiased colors of galaxies. Thus, we used the \texttt{cModelMag} for the total magnitude, and the \texttt{modelMag} for measuring galaxy colors in the following analysis. For this we used SDSS Data Release 16 \citep[DR16;][]{2020ApJS..249....3A}.

We used the Subaru prime focus camera (Supreme-cam) \textit{B}, \textit{V}, and \textit{Rc}-band ($34\arcmin\times27\arcmin$) photometric images from the Subaru public data archive, Subaru-Mitaka Okayama-Kiso Archive System \citep[SMOKA;][]{2002ASPC..281..298B}, to visualize the galaxy distribution of A2261 (Figure \ref{fig:fig1}). We used Subaru images because they have a wider field of view and better quality than SDSS. Figure \ref{fig:fig1} is a color image within the half virial radius of A2261 (the virial radius is 1.114 Mpc, as measured from our own data (see Section \ref{subsec:Comp})), and shows the dominance of the BCG. There is no galaxy that has a similar brightness to that of the BCG (\texttt{cModelMag}$\_$r=15.57) within the half-virial radius region of the cluster even when we see it by eye. The other physical properties of A2261 are summarized in Table \ref{tab:a2261}.

\subsection{Spectroscopic data} \label{subsec:Obs}
\begin{figure*}[htp]
\includegraphics[scale=.36]{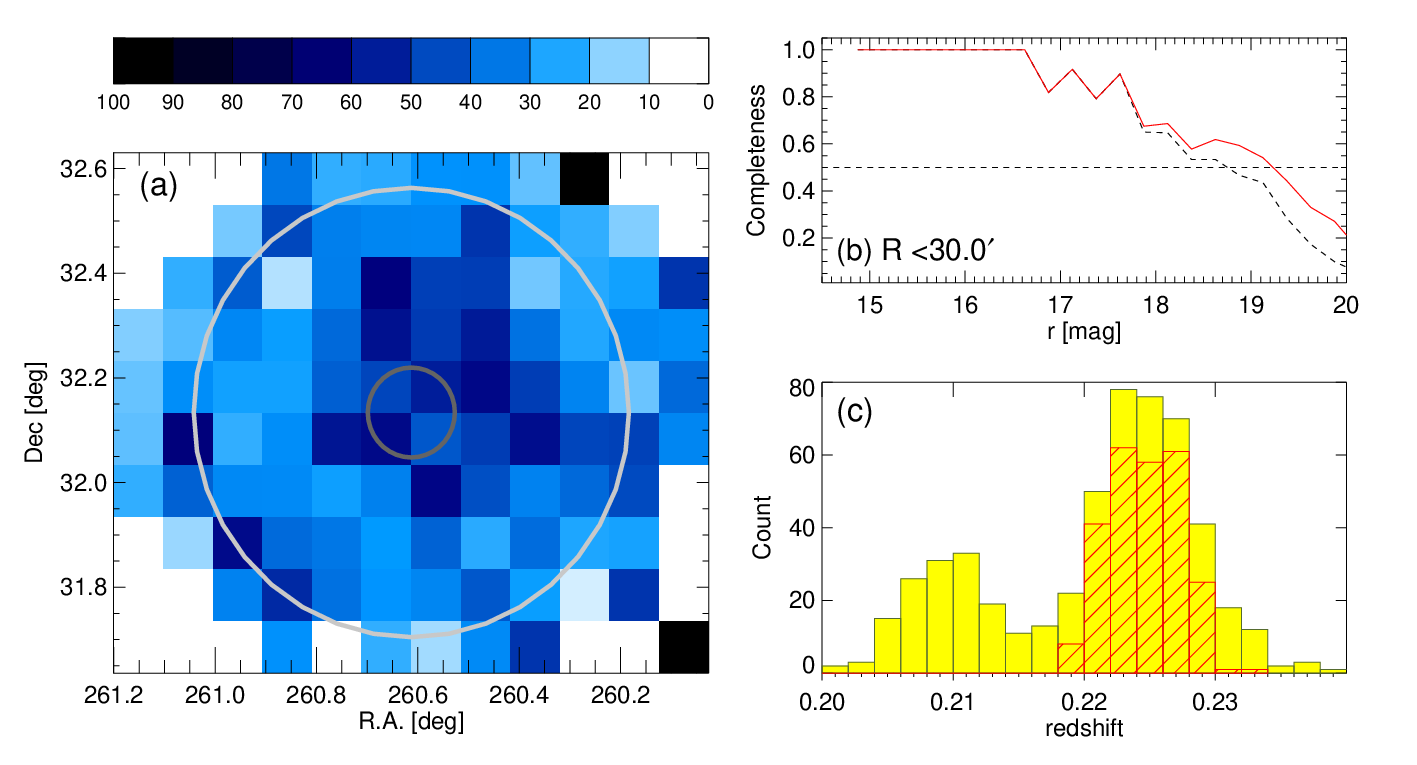}
\caption{\textbf{(a)} Two-dimensional completeness map of total spectroscopy data. Deep gray and mild gray circles show $1R_{200}$ and $5R_{200}$ each.  \textbf{(b)} Spectroscopic data completeness as function of \texttt{cModelMag$\_$r}. Black dashed line is original data, and the red line is observation added completeness. \textbf{(c)} Enlarged redshift distribution of spectroscopic observation data. The yellow-colored histogram shows the total galaxy distribution within redshift of 0.2$\leq z\leq$0.24. The overlapped red hashed histogram indicates the redshift range of membership galaxies from the caustic profile. The caustic method will be described in Section \ref{subsec:Comp}. (a) and (b) used data within 30$\arcmin$ from the cluster center and \textit{r}$<$20. }\label{fig:comp}
\bigskip
\end{figure*}

A2261 is one of the clusters in the Hectospec cluster survey \citep[HeCS;][]{2013ApJ...767...15R} catalog. HeCS has observed 58 galaxy clusters within z= 0.1$\sim$0.3 based on SDSS DR7 \citep{2009ApJS..182..543A}, the ROSAT All-sky survey \citep{1999A&A...349..389V}, and NASA/IPAC Extragalactic Database (NED). The member clusters in HeCS catalog were selected based on X-ray detection, thus most clusters are massive. Candidate cluster member galaxies were selected using the Red-sequence technique \citep{2000AJ....120.2148G}, and so confirmed cluster members are biased towards redder colors and brighter magnitudes among all of the true members. The spectroscopy membership data for A2261 in the HeCS catalog is at z= 0.20$\sim$0.24 and within 1 degree from the cluster center. We obtain 209 spectroscopic redshifts having \textit{r} {$\leq$} 21 from the catalog. This data has an average completeness of $31\%$ within $1R_{200}$ and $20\%$ completeness within $5R_{vir}$.

Because of the low completeness of existing data, we tried to obtain complementary data for A2261 by conducting our own observations with Hectospec \citep{2005PASP..117.1411F} on the MMT in the observing period 2017A (Program ID: MMT-2017A-3). First, potential member galaxies were selected using photometric redshifts \citep{2016MNRAS.460.1371B} from SDSS within a 1 degree F.o.V. Target galaxies were extracted within $z_{cl}\pm 0.0527$, which is from $\Delta z_{phot}=z_{cl}\pm(1+z_{cl})0.043$ (0.043 is the RMS of the \texttt{PhotoZErrclass} of the SDSS), \texttt{modelMag$\_$g} - \texttt{modelMag$\_$r} $<$ 1.7, and \texttt{cModelMag$\_$r}$<$22.5. The color cut and magnitude cut were chosen based on a color-magnitude diagram of the SDSS galaxies with the HeCS confirmed members overlaid. Galaxies in the archive data were excluded from the target list. Target priority was decided based on \textit{r}-band magnitude and distance from the cluster center. Hectospec can place 300 fibers simultaneously and has a 1-degree field of view. We assigned only 250 fibers for target galaxies, because of the fiber collision problem that occurs when the distribution is very dense within a small region. Extra fibers were placed for standard stars and sky measurements. For on-source integral time, we allocated 20 minutes {$\times$} 3 times (1 hour) per configuration to get a S/N $>$ 5 for a blue galaxy, which has \textit{r} {$\leq$} 21.5, and get a S/N $>$ 3 for a red galaxy, which has \textit{r} {$\leq$} 20. Two configurations were executed (2 hours).

The observed data were processed by the \texttt{IDL HSRED v2} package (developed by Richard Cool) using the Hectospec data reduction program. This program provides a list of best-fit redshift values. We tried to estimate the redshift correctly by matching \texttt{RVSAO} best-fit list with visual inspection data that are verified through the IDL based spectrum visualize program, \texttt{Specpro} \citep{2011PASP..123..638M}. In the visual inspection, we usually used H{$\alpha$}, [OIII], H{$\beta$}, and [OII] emission lines, and Ca II H$\&$K, MgII, and G-band absorption lines. We classified 3 kinds of flags for visual inspection : `flag1' for 3 lines matched, `flag2' for 2 lines or uncertain cases, and `flag3' for the zero or one line case. The inclusion of flag 2 objects only provides 19 additional objects and these are less trustworthy. Therefore, we decided to use only flag 1 objects in this study.
After matching the best-fit list with the visual inspection data, we obtained the spectral redshifts of 371 galaxies with a redshift uncertainty of $\sim$40 km s$^{-1}$.

Figure \ref{fig:comp} presents the histogram and completeness of the total data sample of A2261. The completeness is calculated using galaxies having \textit{r}$<$20 and 30$\arcmin$ radius range in the SDSS photometry data. Through our observations, the total completeness is increased (see Figure \ref{fig:comp}-(a)). The mean completeness within $1R_{200}$ is increased to $45\%$ and mean completeness within $5R_{200}$ is $24\%$. The 2D completeness map shows a uniform distribution by decreasing with radius. In Figure \ref{fig:comp}-(b), the completeness as a function of \textit{r}-band magnitude is shown. The completeness reaches 50$\%$ (horizontal dashed line) at 19.3 magnitude with the addition of the new data. Figure \ref{fig:comp}-(c) has a dominant peak between z= 0.218$\sim$0.232, which is similar to the known redshift range of A2261 from \citet{2013ApJ...767...15R}. Dominant background redshift peaks are not visible, but dominant foreground redshift peaks can be seen.

\subsection{Cluster member galaxy identification} \label{subsec:Comp}
\begin{figure*}[htp] 
\includegraphics[scale=.40]{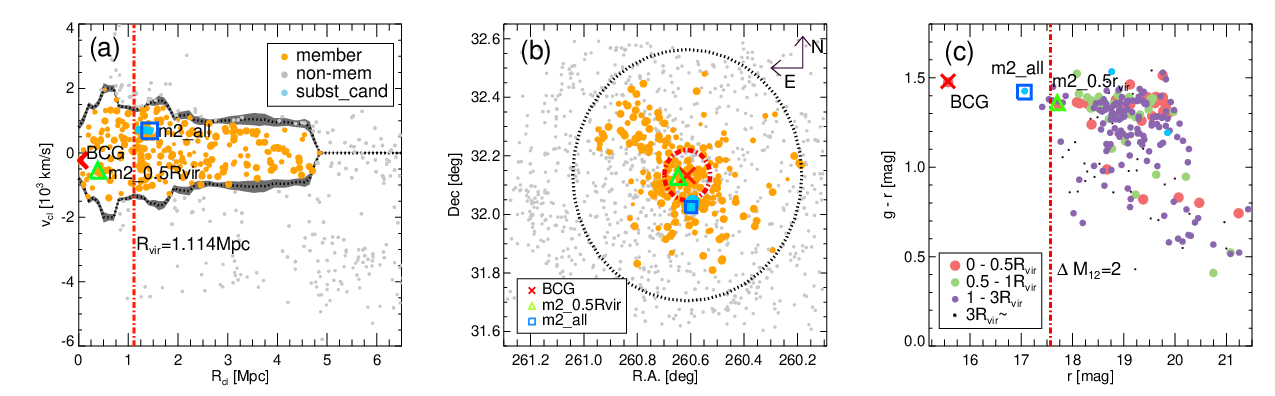}
\caption{\textbf{(a)} Phase-space diagram of A2261. Black dotted line with gray shaded region denote caustic profile and errors. Member and non-member galaxies are represented as orange and gray circles, respectively. Sizes of orange circles depend on the magnitude of galaxies. Red dashed-dotted line shows {$R_{200}$} location. Red cross, green triangle, and blue square located at BCG, second brightest galaxy within the half radius (m2$\_$0.5$R_{vir}$), and the second brightest galaxy at all members (m2$\_$all), respectively. Mild-blue circles are group candidates, as discussed in Section \ref{sec:rean} \textbf{(b)} Spatial distribution of the entire data set. The symbols and color-coding are same as in (a). The black-dashed line shows $5R_{vir}$. \textbf{(c)} Color-magnitude diagram of spectroscopic members of A2261. X-axis is \texttt{cModelMag$\_$r}, and Y-axis is \texttt{modelMag$\_$g} - \texttt{modelMag$\_$r} color from the SDSS DR16. The pink, mild green, and cyan circles represent the member galaxies within $0\sim0.5, 0.5\sim1$, and $1\sim3 R_{vir}$ bins, respectively. Black dots are member galaxies beyond $3R_{vir}$. Other symbols are the same as in panel (b). The vertical red dashed-dotted line is two magnitudes lower than the BCG's magnitude.}\label{fig:caustic}
\end{figure*}

In this paper, we use the Caustic method \citep{1997ApJ...481..633D, 2013ApJ...768..116S} to obtain the member classification and to calculate new physical parameters based on the new membership. The caustic method \citep{1997ApJ...481..633D} identifies cluster members using the escape velocity of the cluster. The escape velocity lines from the caustic method have a trumpet-like shape. The caustic method is less severely affected by the dynamical equilibrium than traditional techniques \citep[see][]{2022MNRAS.509.3470M}, thus it is possible to find unrelaxed first-infaller cluster members even outside of the virial radius of the cluster. However, its success relies on having enough number of galaxies in the sample ($\sim 200$) in order to detect the caustic \citep{2013ApJ...768..116S}, a number that we easily exceed with our new spectroscopy.

We used the \texttt{Caustic App v1.6} free software tool, developed by \citet{2013ApJ...768..116S}\footnote{Software was obtained by private communication.}. We set it to calculate a new center position and properties of the cluster. It identified 257 galaxies as members from a total of 526 galaxies. Figure \ref{fig:caustic}-(a) shows identified member galaxies of A2261 (orange-colored circles) and indicates the calculated caustic line with error range (dashed line and gray shade) obtained from the \texttt{Caustic APP}.

\textbf{New properties:} We found that member galaxies are distributed out to $\sim$5Mpc from the cluster center with a cluster velocity dispersion of $\sim$661km s$^{-1}$. The new virial radius of the cluster was calculated as $R_{200}$=$\sim$1.114Mpc. Both velocity dispersion and $R_{200}$ are calculated using member galaxies that are inside the caustic profile, and therefore can be found out to 5 $R_{200}$.

The new center point had a negligible change (RA:260.6129,  DEC:32.1338) from HeCS data. This result is similar to the X-ray center position in ACCEPT \citep{2009ApJS..182...12C} (RA: 260.6135, DEC: 32.1329) and BCG position (RA: 260.6132, DEC: 32.1325). The projected positional offset between the new center point and the X-ray center is 13 Kpc (0.012 $R_{vir}$), and the projected offset between the new center point and the BCG is 17 Kpc (0.015 $R_{vir}$). The radial velocity offset between the BCG and the new center point is 245 km s$^{-1}$ (0.37$\sigma_{cl}$). These offset values show that center point of cluster and BCG are well matched, which can be considered an indicator of a cluster with a relaxed dynamical state, but radial velocity offset shows quite a difference.

\textbf{Main features:} Figure \ref{fig:caustic}-(b) shows the spatial distribution of the member galaxies, non-member galaxies, BCG, second brightest galaxy within half virial radius(m2$\_$0.5$R_{vir}$) and whole cluster member(m2$\_$all), and group candidate positions (all are discussed with more detail in Section \ref{subsec:ds}). The second brightest galaxy within half virial radius(m2$\_$0.5$R_{vir}$), second brightest galaxy with whole cluster member(m2$\_$all) and group candidate are highlighted on the color-magnitude diagram of the members (see Figure \ref{fig:caustic}-(c)). The entire shape of the spatial distribution of the member galaxies is elongated along the North-East direction (towards the upper-left corner in panel (b)). Although we observe a 1-degree field of view with no directional bias (see Figure \ref{fig:caustic}-(b)), member galaxies (as identified by the caustic method) were found all the way out to 5 $R_{vir}$ in this direction. This is far beyond the typical radius reached by back-splash galaxies \citep{2004ogci.conf..513M, 2011MNRAS.411.2637P}. This extended feature has a similar velocity dispersion as the cluster value. We suspect this is evidence of a nearby filament that is connected with A2261, and we find further evidence for this picture in Section \ref{subsec:lss}.

\textbf{Color-magnitude diagram:} The color-magnitude diagram (CMD) is also shown in panel (c). To make this diagram, the SDSS model g and r magnitude are used for our spectroscopically confirmed members. We limit our sample to galaxies out to 3$R_{vir}$ from the cluster center to exclude possible filament galaxies, but not neglect possible back-splash galaxies. In the conventional definition of a fossil system, the second brightest galaxy must be within 0.5$R_{vir}$ and at least two magnitudes fainter than the BCG. As shown in Figure \ref{fig:caustic}-(c), the pink colored circles, which represent galaxies inside 0.5$R_{vir}$, are all two magnitudes fainter, meaning A2261 should indeed be classified as a fossil system. We label the brightest galaxy among them as m2$\_$0.5$R_{vir}$, as it is the second brightest galaxy to the BCG inside of 0.5$R_{vir}$ (indicated with a green triangle symbol).

Besides, we can also find another bright galaxy. This is located just beyond the virial radius of the cluster (at $\sim1.27 R_{vir}$ $(\sim1.41 Mpc)$) and is actually brighter than m2$\_$0.5$R_{vir}$, and has a magnitude gap of less than two magnitudes ($\Delta M_{12}=1.49$) with the BCG. This is less than the fossil magnitude gap criteria (Figure \ref{fig:caustic}-(c)) but because it is beyond the virial radius, it does not break the fossil definition. Because this galaxy is the second brightest galaxy compared to the BCG of all of our cluster members, we hereafter label it m2$\_$all. We investigate its features in more detail in Section \ref{subsec:xray}, also.

\section{Results} \label{sec:rean}

Using our new membership galaxy data, we wish to measure the dynamical state throughout the cluster, from its core to its outskirts. First, we check the values of dynamical state indicators using our spectroscopic data in Section \ref{subsec:otherpara}. In Section \ref{subsec:ds}, we use the D-S test to search for substructures. We also compare with X-ray emission contours in Section \ref{subsec:xray}. The probability of the presented group candidate is checked in Section \ref{subsec:prob} and the presence of a nearby large-scale structure is confirmed in Section \ref{subsec:lss}.

\subsection{Dynamical indicators \label{subsec:otherpara}}
\begin{figure*}[htp]
\begin{center}
\includegraphics[scale=.85]{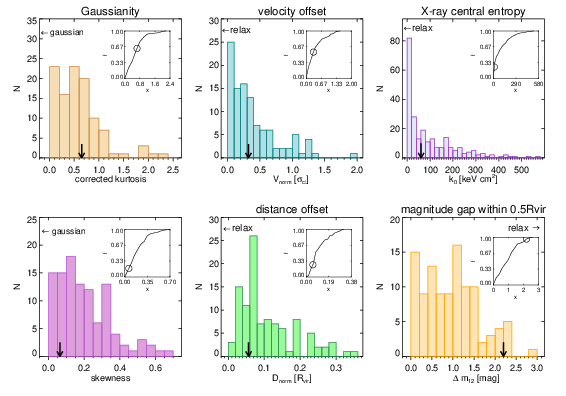}
\caption{Histograms of dynamical state indicators. The values of each parameter for A2261 are shown as a black vertical arrow (also provided in table \ref{tab:indicator}). Horizontal arrows give the direction in which a cluster with a more relaxed (or gaussian) dynamical state would be found. Inset panels show the normalised cumulative fraction. X axis is each value and Y axis is cumulative fraction. The circle symbol indicates A2261's location in the distribution.}\label{fig:histo}
\end{center}
\end{figure*}

Here, we measure various dynamical state indicators for A2261, as listed individually below. We compare these measurements to the same measurements for a comparison sample of clusters. To build the comparison sample, we begin with 212 spectroscopically observed galaxy clusters from CIRS  \cite[Cluster Infall Regions in SDSS,][72]{2006AJ....132.1275R}, HeCS (Hectospec Cluster Survey, 58), HecS$\_$SZ \cite[Hectospec Cluster Survey of SZ-selected clusters,][53]{2016ApJ...819...63R}, HeCS$\_$red \cite[Hectospec Cluster Survey of red-sequence-selected clusters,][23]{2018ApJ...862..172R} project catalogs , and etc \cite[][6]{2014ApJ...797..106H} clusters. Those include massive and X-ray detected clusters within z= 0.001 $\sim$ 0.29. Among them, we choose 106 galaxy clusters for our comparison sample that have up to 70$\%$ spectroscopic completeness. The minimum, maximum, and average number of the member galaxy of 106 clusters are 16, 1349, and 205, respectively. Figure \ref{fig:histo} shows histograms of each type of dynamical state indicator of the 106 galaxy clusters. Inset panels in each histogram are the normalized cumulative fraction of each indicator. In this way, we can compare the dynamical state of A2261 (shown with a black arrow) to that of the comparison sample, measured in the same way. 

\textbf{Gaussianity:} The kurtosis and skewness are used to check the gaussianity of the velocity distribution of cluster members \citep{1996ApJS..104....1P, 2007ApJ...662..236H, 2012A&A...540A.123E}. The kurtosis values in this figure are given as the difference with respect to the Gaussian value of 3. Thus, the direction along the x-axis towards a value of zero means a more gaussian distribution. The gaussianity of the velocity distribution of the cluster can be interpreted as a measure of the orbital relaxation of member galaxies, as large numbers of first infallers could make the distribution increasingly non-gaussian. Also, if there is a significant substructure, or the cluster is merging, non-gaussianity would increase further. 

The calculated values of skewness and kurtosis are 0.056 and -0.63, respectively. This means that A2261 is a little left-skewed and has a slightly peaked central distribution. We also see the location of the value of A2261 in histogram of sample galaxy clusters and do a Shapiro-Wilk normality test \citep{10.1093/biomet/52.3-4.591} that is usually proper for samples of a small size. We get a p-value from the Shapiro-Wilk test of $P_{SW}\simeq$0.044, which indicates that A2261 has a velocity distribution that deviates from a normal distribution (i.e., it is not Gaussian in shape).

\begin{table}[htp]
    \caption{Values of dynamical state indicators in Figure \ref{fig:histo}.}\label{tab:indicator}
\begin{center}
    \begin{tabular}{ccc } 
    \hline
    Dynamical & A2261 & criterion\\[0.5ex]
    Indicator & value & value\\[0.5ex]
     \hline
     Corrected Kurtosis & -0.63 & 0\\[0.5ex]
    
     Skewness & 0.056 & 0\\[0.5ex]
   
    $P_{SW}$ & 0.044 & $>$ 0.05\\[0.5ex]
    
    Velocity Offset [$\sigma_{vir}$] & -0.32 & $<$ $\vert0.1\vert$\\[0.5ex]
    
    Distance Offset [$R_{vir}$]& 0.056 & $<$ $\vert0.1\vert$ \\[0.5ex]
    
    ${\Delta M_{12}}$ & 2.2 & $>$ 2\\[0.5ex]
    
    $K_0$ [KeV $cm^2$] & 61.08  & 30\\[0.5ex]
    \hline
    \end{tabular}
\tablecomments{ $P_{SW}$ is the p-value of the Shapiro-Wilk normality test. The velocity offset is normalized by the velocity dispersion and the distance offset is normalized by the virial radius of A2261. ${K_0}$ is from the ACCEPT project \citep{2009ApJS..182...12C}. Specific explanations for indicators are in Section \ref{subsec:otherpara}.}
\end{center}
\end{table}

\textbf{Offsets:} Offset indicators have been frequently used as dynamical state indicators in the past, with small offsets tending to indicate a more relaxed state. However, the value of measured offset parameters can be susceptible to projection effects. There are various kinds of offset parameters, and some are measured from the luminosity-weighted center or X-ray center \citep{2008ApJS..174..117M,2016ApJ...824..140R,2017ApJ...836..105K} of the cluster. 
In this section, we consider two offset parameters. The distance offset is the calculated difference between the luminosity-weighted center of the cluster and the BCG. For the luminosity weighting, we use the \textit{Petrosian r} magnitude from SDSS DR16 that is provided by the above catalogs (there is little difference if we use \texttt{cModelMag} instead). When we calculate luminosity weights, the BCG magnitude is excluded from the calculation. Thus, the luminosity weighted center represents the center among satellite member galaxies. The distance offset value is normalized by $R_{vir}$ ($D_{norm}$). The velocity offset is measured in a similar way, only considering velocity differences instead of position differences (i.e., comparing the BCG velocity to the luminosity weighted mean velocity of the satellite members) and normalized by $\sigma_{vir}$ of each cluster ($V_{norm}$). There is not an exact value for dividing relaxed and unrelaxed clusters, but there are many studies that demonstrate that small offset means a more relaxed state \citep{2011A&A...526A.105Z, 2019ApJ...887..264R}. If we assume that a value of less than 10$\%$ of the normalized value ($<\vert0.1\vert$) would be a relaxed system, then the distance offset of A2261 is a relaxed value, but velocity offset of A2261 is a more moderate value.

\textbf{Magnitude gap and X-ray central entropy:} The magnitude gap is calculated using galaxies within 0.5$R_{200}$ after double-checking the BCG in the CMD of each cluster. As we mentioned before, conventionally $\Delta M_{12} > 2$ is considered as a relaxed state \citep[e.g.][]{2003MNRAS.343..627J}.
The X-ray central entropy, $K_0$ values are from 241 galaxy clusters in ACCEPT \citep{2009ApJS..182...12C}, and normally lower than 30KeV $cm^{2}$ is considered a relaxed state \citep[e.g.][]{2018MNRAS.481.1809B}. For A2261, the magnitude gap is larger than the criterion and the X-ray central entropy has a moderate value. 

Considering the above indicators, most of the values of A2261 are biased towards relaxed (or Gaussian) states. The individual values of A2261 about dynamical indicators can be found in Table \ref{tab:indicator}.

\subsection{D-S test} \label{subsec:ds}
\begin{figure*}[htp]
\begin{center}
\includegraphics[scale=.60]{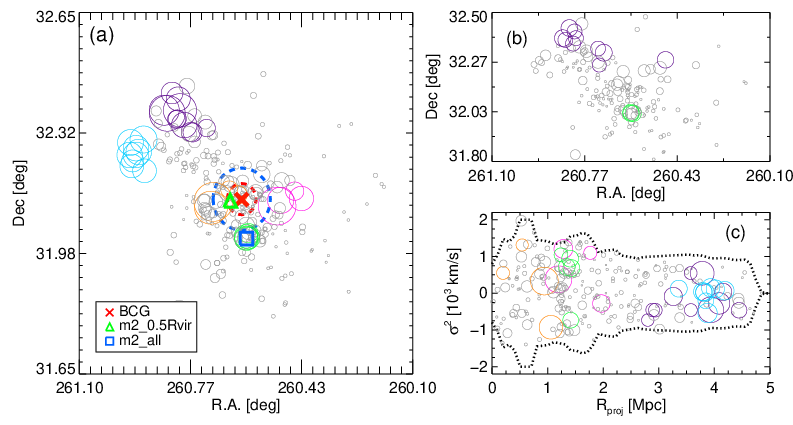}
\caption{ \textbf{(a)}Bubble plot of A2261 with all member galaxies from new data. The gray circles (bubbles) show the velocity deviation between the sub-cluster candidate and the entire cluster. The size of circles is proportional to $\delta^2$. Other colored circles show large deviation groups ($\delta^2 > 2$). Red crosses, green triangles, and blue squares indicate the BCG, second brightest (m2$\_$0.5$R_{vir}$) galaxy within the half virial radius, and second brightest (m2$\_$all) galaxy of all member locations, respectively. 0.5$R_{vir}$ is shown as a red dashed line circle, and 1$R_{vir}$ is drawn by a blue-dashed line circle. \textbf{(b)} The result of D-S test using previous data. Color-coding is the same with (a). \textbf{(c)} Phase-space distribution of large deviation groups. Dashed-line shows the caustic profile. Color-coding is same with (a).} \label{fig:bubble}
\end{center}
\end{figure*}
\begin{figure}[htp!]
\begin{center}
\includegraphics[scale=.65]{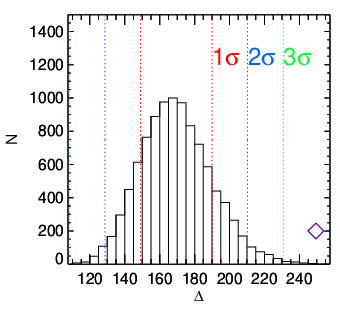}
\caption{Histogram of the Delta ($\Delta$, see formula 4) values after randomly shuffling their line-of-sight velocities 100 thousand times. The red, blue, and green dotted-line show the 1, 2, and 3$\sigma$ value, respectively. The purple diamond symbol shows the measured $\Delta$ value for our A2261 data set. It is located at a 3$\sigma$ higher value than a random distribution indicating that A2261 distribution differs significantly from a random distribution, and contains substructures.} \label{fig:his}
\end{center}
\end{figure}

The existence of substructure is often used as a proxy of the unrelaxed state of the cluster \citep[see][]{2008ApJ...682L..73S,2010MNRAS.409..169S}. There are many methods to find substructures inside a cluster. One of the methods, the Dressler$-$Shectman test \citep[][D-S test]{1988AJ.....95..985D} is applied to our A2261 spectroscopic sample in this study. 

To identify substructures, the D-S test compares the differences of the mean velocity and velocity dispersion between a global value (using all cluster galaxies) and a local value (specified number of its nearest galaxies).

First, we calculate the mean velocity and velocity dispersion for all cluster members, which is represented as $\bar{v}_{cl}$ and $\sigma_{cl}$.

Then, we calculate the mean velocity ($\bar{v}^{i}_{local}$) and velocity dispersion ($\sigma^{i}_{local}$) of sub-cluster candidate galaxies, which are the neighbor galaxies around the $i$-th galaxy, following 

\begin{equation}
\bar{v}^{i}_{local} = \sum\limits_{j=1}^{N_{nn}} \frac{1}{N_{nn}} \frac{c (z_{ij}-z_{cl})}{1+z_{cl}}  ,
\end{equation}

\begin{equation}
\sigma_{local}^{i} = \sum\limits_{j=1}^{N_{nn}} \frac{1}{N_{nn}} \frac{(cz_{ij}-\bar{v}_{local}^i)^2}{1+z_{cl}}
\end{equation}

where $z_{ij}$ is the redshift of $j$-th closest galaxy to $i$-th galaxy ($i$ for all cluster member, $j$ for sub-cluster galaxies). $z_{i1}$ is the $i$-th galaxy itself. $N$ is the number of member galaxy, and $N_{nn}$ is the number of candidate member galaxies for a sub-cluster. $N_{nn}=\sqrt{N}$ is adapted from \citet{1996ApJS..104....1P}. For A2261, $N_{nn}=16$ is selected. We also tested $N_{nn}=11$, which is the original number from \citet{1988AJ.....95..985D}, but it did not significantly change our results.

Finally, the deviation ($\delta^2$) of each local value from the global value is defined by 
\begin{equation}
\delta^{2}_{i} = (\frac{N_{nn}}{\sigma_{cl}^{2}})[(\bar{v}^{i}_{local}-\bar{v}_{cl})^{2}+(\sigma^{i}_{local}-\sigma_{cl})^{2}] .
\end{equation}

If the mean velocity and the velocity dispersion of the sub-cluster candidate are highly different from the cluster, the deviation value increases. We visually represent the calculated deviations with a circle whose size is proportional to $e^{\delta_{i}}$, in Figure \ref{fig:bubble}, which is known as a \textit{`bubble plot'}.

If galaxies are gravitationally bound to each other and form a substructure, those galaxies should have a similar velocity and velocity dispersion with other galaxies nearby to their location. Therefore, large aperture circles located close together or overlapping in the bubble plot can be considered as probable substructures. 

\textbf{Specific membership:} One of the weaknesses of the D-S test is that it can not distinguish which specific galaxy is a group member or not. In the literature, there are some attempts to roughly define membership of individual galaxies using $\delta$ values as a threshold \citep{2009A&A...495..379B, 2013MNRAS.431.2111J}. A value of $\delta^2>2$ has been used to approximately define the location of groups.

Applying this criterion to our results, colored circles in Figure \ref{fig:bubble}-(a) show the probable locations of substructures. Each separated substructure is colored differently. Near the $\sim1R_{vir}$ location, there are substructures in the east (orange circles), west (pink circles), and south (green circles). In the phase-space diagram, the east (orange) and west (pink) circles have different velocities compared with nearby galaxies, and are located far away from each other. Those are less likely to be genuine substructures, and are perhaps just projection effects. However, the southern-side green circles are located nearby to each other and have a similar size. 

We also double-checked the southern-side green colored circles in the phase-space diagram (see Figure \ref{fig:bubble}-(c)), spatial distribution and CMD (see Figure \ref{fig:caustic}), and indicated it as `group candidate'. 

On the North-East side, there are two groups of colored circles that show interesting features. The upper circles (purple) have a scattered distribution, but some of the lower circles (cyan) are well aligned and overlapped in the phase-space diagram (Figure \ref{fig:bubble}-(c)). This group-like feature was not detected in the previous archive data, and is only revealed when we add our new observation data (see Figure \ref{fig:bubble}-(b)). We also tested with a different number of local group membership, but this substructure remains visible. This likely suggests that we have detected a group of galaxies in the cluster outskirts near 4$R_{vir}$ that is being fed into the cluster along a connected filament.

\textbf{Cumulative deviation ($\Delta$):} As another test for the existence of substructure, we compare the cumulative deviation, $\Delta$ values of A2261 to the cumulative deviation distribution of random distributions that have equal sample size. $\Delta$ is defined by

\begin{equation}
\Delta = \sum\limits_{i=1}^N \delta_{i},
\end{equation}

Where N is the total number of member galaxies that are found beneath the caustic (see Section \ref{subsec:Comp}).

We compare this with randomly generated cumulative deviations. We generate 100,000 random distribution data sets by shuffling the velocities of the original data. Then we calculate the p-value of the observed value to the randomly generated sets (see Figure \ref{fig:his}).

\begin{equation}
P = \sum(\Delta_{random}>\Delta_{original})/N_{random},
\end{equation}

Where $N_{random}$ is 100,000. $\Delta_{random}$ and $\Delta_{original}$ are measured from the random distribution (made by shuffling the velocities) and the original distribution of A2261, respectively. 
The mean value of random sets ($\Delta_{random}$) is 169.3 with a standard deviation of 20.4. The original cumulative deviation from the original distribution ($\Delta_{original}$) is 249.4, which is 3$\sigma$ above. Thus, the observed cumulative deviation is 3$\sigma$ above the mean value of the randomly generated sets (a p-value $\leq$ 0.01). This means the original distribution is unlikely to be a random distribution with a high significance. Physically, this can be interpreted as meaning that substructure exists in the A2261 members with a high probability, which could be considered as a feature of an unrelaxed cluster. If we exclude galaxies beyond 1 $R_{vir}$, the original distribution is 2 $\sigma$ above.

\subsection{The group candidate in X-rays \label{subsec:xray}}

\begin{figure*}[htp!]
\begin{center}
\includegraphics[scale=0.41]{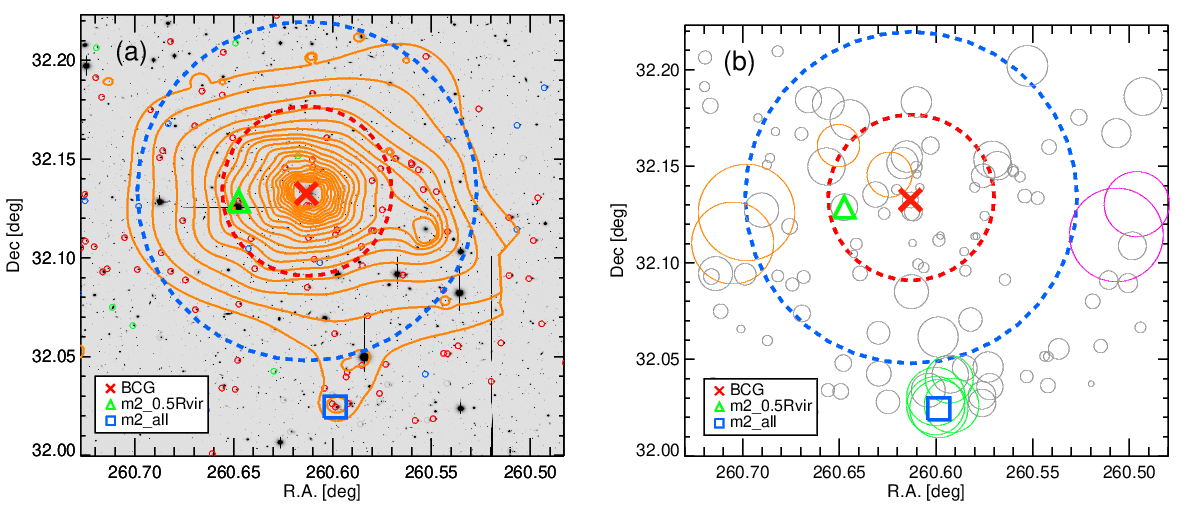}
\caption{\textbf{(a)} Subaru \textit{Rc}-band image of A2261. Orange colored lines are X-ray emission contours from Chandra observations \citep{2008ApJS..174..117M}. Blue and red dashed circles show 1 and 0.5 virial radii, respectively. Red crosses, green diamonds and blue rectangular symbols indicate the positions of the BCG, original second brightest (m2$\_$0.5$R_{vir}$), and real second brightest galaxy (m2$\_$all). Small circles with red, blue, and green colors are absorption, emission, and weak emission line detected member galaxies, respectively. \textbf{(b)} D-S test bubble plot with the same coverage as in Figure (a).}\label{fig:xray}
\end{center}
\end{figure*}

To follow up on the possible group candidates from D-S test, we also studied the X-ray observations of A2261 as well. We used the X-ray surface brightness contour profile from \citet{2008ApJS..174..117M}. This X-ray data is from the Chandra Advanced CCD Imaging Spectrometer (ACIS) observation (OBJID:5007) in 2004. The contour map is available on their website\footnote{http://cxc.cfa.harvard.edu/cda/Contrib/2007/MAUG1}. Figure \ref{fig:xray} shows an image with the smoothed X-ray emission contour and our spectroscopic member galaxies on the Subaru \textit{Rc}-band optical image of A2261. The strength of the X-ray emission decreases with radius, and the overall shapes are well-ordered.

However, there are a few clumps visible in the contours. \citet{2019A&A...622A..24S} tested whether the west-side clump might be an infalling group, but they rule this out. At its location, they find a background red galaxy at z=0.236, slightly beyond the cluster membership redshift range ($0.2165< z_{mem} <0.2321$). There is also a foreground galaxy \cite[photometric redshift =0.088,][]{ 2007AN....328..852C}, and a star, all located very close to each other in the sky. But there is neither a galaxy group nor a host of a hot gas clump such as AGN, QSO, or radio galaxy. The spectra of the galaxies also do not show any significant broad emission lines for AGN, QSO, or radio galaxy.

Only the X-ray clump located towards the south matches with the spectroscopically confirmed member location. There are 5 member galaxies with absorption line features. Among nearby galaxies, 3 galaxies are matched with the location of the high deviation assembled circles at the D-S test (Figure \ref{fig:xray}-(b)), and group candidate location on the caustic profile (Figure \ref{fig:caustic}-(a)). This provides further confirmation for the presence of a real substructure at the south of A2261. Furthermore, the tail-like shape of the x-ray contours centered on the group gives the impression that the group is interacting with the cluster's ICM. The direction of the x-ray tail points roughly inwards as if the group is exiting the cluster, although we cannot be certain of the influence of projection effects.

Interestingly, one of the members in the X-ray clump is m2$\_$all, which has a brighter magnitude value than m2$\_$0.5$R_{vir}$. m2$\_$all is sufficiently bright that it would break the fossil definition if it were inside the half virial radius ($m_{bcg}-m2\_all=1.49$), but it is currently found beyond the virial radius. Therefore, assuming the group previously crossed the cluster in the past, here we attempt to estimate timescales for the cluster to transition from its fossil status. To do this, we must make many simplifying assumptions.

We do not know the true orbital velocity of the group candidate. We can only measure its line-of-sight velocity. If we assume the average velocity of the group candidate is similar to the 3D cluster velocity dispersion, and that the cluster has an isotropic velocity distribution, then its 3D velocity is $\sqrt3\times 661km/s$=1145km/s = 1.145 Mpc/Gyr. Now, we estimate how long it would take for the group to cross from entering the half virial radius of the cluster on the opposite side, until reaching the half virial radius on its current side of the cluster. We find this would take about $\sim$0.97 Gyr in total. During this time, A2261 would cease to be a fossil system, only transitioning back to being a fossil when the group leaves the half virial radius on the South-West side.

We now estimate how long it would take, after leaving the half virial radius, to reach its currently observed position. This is the amount of time that the cluster has been a fossil system since the transition. We do not know the true separation between cluster and group (in three dimensions). If we assume spherical symmetry for the group projected distance, the average expected 3D separation between the group and the cluster center is $\pi/2\times1.35 Mpc$ = 2.12 Mpc. The time required for the group to go from 2.12 Mpc to half the virial radius is (2.12-0.55) Mpc/(1.145Mpc/Gyr) = 1.36 Gyr, or, if the projected distance is the true distance, 0.8 Gyr.

Under this assumption, the timescales mean that the cluster will remain as a fossil for another 0.8 Gyr (assuming the true separation distance is equal to the projected distance) or $\sim$1.36 Gyr (assuming the true separation distance is 3D distance), at which point it will enter the half virial radius and break the fossil definition.

In summary, we estimate that the presence of the group containing m2$\_$all likely causes A2261 to transition between fossil and non-fossil status on timescales of roughly a gigayear.

\subsection{Cluster member probability of group candidate \label{subsec:prob}}

\begin{figure*}[htp]
\begin{center}
\includegraphics[scale=.35]{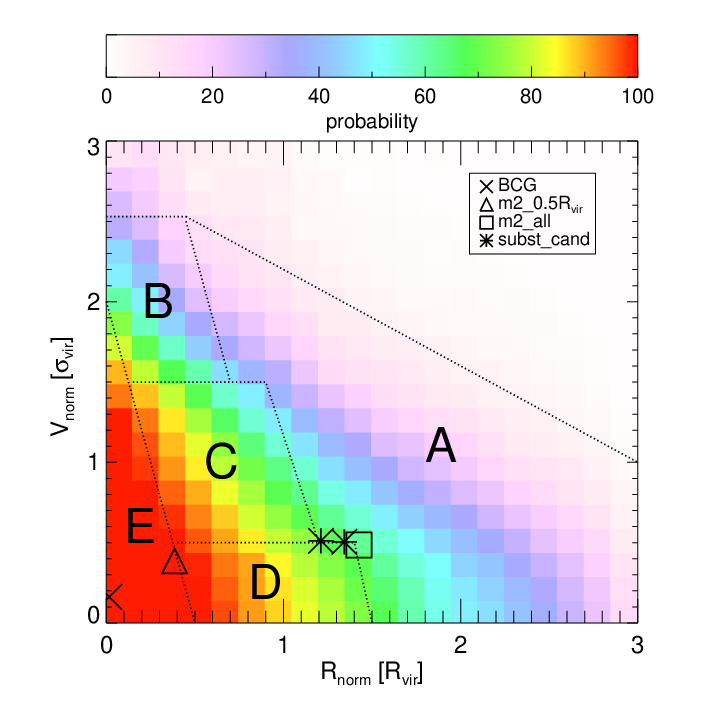}
\caption{Phase-space membership probability plot using Illustris-TNG simulation. Probability is colored as follows the color bar of the top. Each symbol is represented in the legend. X and Y axis are normalized by cluster virial radius and velocity dispersion, each. A to E regions are from Figure 6 in \citet{2017ApJ...843..128R}}\label{fig:prob}
\end{center}
\end{figure*}

Projection effects can make a target seems closer to the cluster than it really is. Thus, we try to apply projected phase-space diagrams using dark matter halos near massive host halos from the Illustris-TNG simulation data \citep{2018MNRAS.475..676S,2018MNRAS.475..648P}. 

\textbf{Membership probability plot within 3$R_{vir}$:} We chose 1566 massive host halos having $M_{vir} \geq 10^{13}M_\sun$ and all subhalos having $M_{vir} \geq 10^{11}M_\sun$. The projected distance and velocity of all halo are normalized by virial radius and the velocity dispersion of each nearby cluster halo in the projected phase-space plane. We calculate the probability that a halo is truly within $3R_{vir}$ of the cluster ($p_{3r}$) at each location in the projected phase-space. This is calculated by taking the number of halos within $3R_{vir}$ ($n_{3r}$) divided by the total number of galaxies ($n_{total}$) including interlopers within specific pixels of projected R-V space.

\begin{equation}
p_{3r} = n_{3r}/n_{total}
\end{equation}

As we can see in Figure \ref{fig:prob}, low to high probability is represented by the colors (see color bar). At the BCG, the probability is nearly 100$\%$ member, m2\_$0.5R_{vir}$ is $\sim$90$\%$. The $m2\_all$ and substructure candidate are at 60$\sim$70$\%$ location. This demonstrates that the substructure detected to the south of the cluster has a high probability to be genuinely near to the cluster from its phase-space location alone. The fact that its X-ray emission appears to be interacting with the cluster strengthens that the substructure is located in the cluster outskirts, and not simply projected onto the cluster. 

\textbf{Regions in phase-space:} We also try to interpret their phase-space locations using the approach of splitting up the regions, as shown in \citet{2017ApJ...843..128R}. In figure 6 of that study, which is a schematic distribution diagram (without error), they divide galaxies into 4 types by their `\textit{time since infall} (t$_{inf}$)', and look at the breakdown of these types in each region of a projected phase-space diagram (or R-V diagram). As galaxies fall into the cluster from outside, they typically tend to follow a track that is from region A through to region E in alphabetical order. As a result, `\textit{first infallers}' dominate in region A, `\textit{recent infallers}' (0$<$t$_{inf}<\sim3.63Gyr$) dominate in region B and C , `\textit{intermediate infallers}' (3.63$<$t$_{inf}<\sim6.45Gyr$) dominate in region D, and `\textit{ancient infallers}' (6.45$<$t$_{inf}<\sim13.7Gyr$) dominate in region E.

The BCG and m2\_$0.5R_{vir}$ fall in region E, which is the region most dominated by `\textit{ancient infallers}'. The $m2\_all$ and substructure candidates both fall on the border of region D, which is the region most dominated by `\textit{intermediate infallers}'.

If there was a minor merger between the cluster and the group that includes $m2\_all$, this amount of time would be enough to relax the optical parameter change (e.g., the magnitude gap \citealt{2010MNRAS.405.1873D, 2017ApJ...845...45K}) of the cluster, but it is likely not enough time for the disturbed X-ray parameters to cool down \citep{2008MNRAS.387..631E}. 

The results from previous sections and this result point to a simple scenario. A2261 may have experienced a past minor merger with a group, and the group has since passed through the cluster (having fallen in at least 3.63 Gyrs ago), and it is now located in the back-splash region.

\subsection{Nearby Large-scale structure environment around A2261 \label{subsec:lss}}

\begin{figure*}[htp]
\begin{center}
\includegraphics[scale=.65]{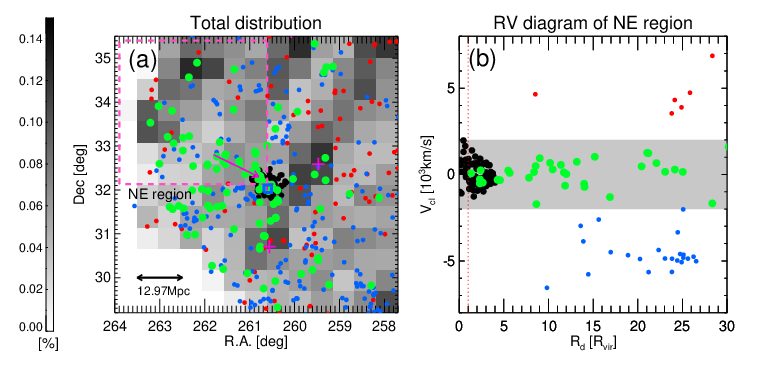}
\caption{\textbf{(a)} Nearby large-scale structure in a $6\arcdeg\times 6\arcdeg$ range (roughly $\sim$60$Mpc^2$ scale) about A2261. The greyscaled color bar shows completeness. Background mosaics show the SDSS spectroscopic completeness on the sky plane. The red, green, and blue colored circles are galaxies within 8000km s$^{-1}\gtrsim$v$\gtrsim$2000km s$^{-1}$, 2000km s$^{-1}\gtrsim$v$\gtrsim$-2000km s$^{-1}$ and -2000km s$^{-1}\gtrsim$v$\gtrsim$-8000km s$^{-1}$, respectively (based on SDSS spectroscopy). Black circles indicate the cluster members from our main spectroscopic data sample. A pink arrow indicates the most probable direction of a filament that connects with the cluster. A blue rectangular symbol shows the location of the group candidate. The pink box is named the North-East region (NE region). Pink crosses are cluster locations around A2261. \textbf{(b)} Phase-space diagram of the NE region. The red dashed line shows the $\sim$1R$_{vir}$ region. The shaded region is $\pm\sim$2000km s$^{-1}$ of the cluster radial velocity. All color coding is same with Figure \ref{fig:lss}-(a).}\label{fig:lss}
\end{center}
\end{figure*}

From previous sections, we can expect that the group candidate has previously penetrated the intracluster medium of A2261. Furthermore, the group candidate is likely located in the cluster outskirts and is not a virialized member of the cluster now. The member galaxy distribution is stretched to the North-East side of A2261, if a filament was connected to the cluster there, and there is evidence for a group infalling within the filament (or a minor merger). In this section, we consider the structure on larger scales around the cluster. 

In the total distribution in Figure \ref{fig:lss}-(a), the background completeness of SDSS DR16 spectroscopy observation shows diagonal strips (running from bottom-left to top-right) due to the manner in which the sky was mapped. If we focus on green circles, which show galaxies with similar velocities to the cluster, we can see a possible filamentary feature. It emerges from the extended feature of a cluster and continues towards the upper-left side, following the direction indicated by the black arrow.

We note that the completeness of the area, where the green circles appear to align along a possible filament, is not higher in that direction. The possible filament lies perpendicular to the enhanced completeness diagonal strip features. This means the possible filament feature is unlikely to be an effect of varying completeness. In panel (b), the R-V diagram of the North-East region (NE region) shows that the galaxies in the possible filament (as seen on the sky) have similar velocities and lie roughly on the plane of the sky over a large range of radius. Most of the galaxies in the NE region are located within $\pm$ 2,000km s$^{-1}$ velocity dispersion, and the feature seems to be directly connected with the cluster on the NE side both spatially and in velocity space too. 

Furthermore, as we saw previously in Figure \ref{fig:caustic}-(b), the location, where the possible filament reaches the cluster, is the same as the location of the extended feature seen in our own A2261 data. This is further support for a filament that connects with the cluster on the NE side. We see only a small gradual velocity increase with increasing distance along the filament as shown.

\textbf{Environment:} We also looked for the presence of previously known groups and clusters in the vicinity of A2261. Fossil systems are thought to live in sparse environments if passive evolution is behind the origin of the high magnitude gap. 

Pink cross symbols show nearby cluster locations from the MCXC catalog \citep{2011A&A...534A.109P}, but all are farther than $\pm$ 8,000km s$^{-1}$ recessional velocity from A2261. Unfortunately, MCXC clusters have only a larger mass than average, because of the limitation of X-ray observation, and so the presence of group scale objects can not be ruled out. Although it appears that there are no cluster-mass neighbors within the $6\arcdeg\times6\arcdeg$ range, our results suggest there are nearby groups like the A2261 group candidates. For example, in Figure \ref{fig:comp}-(c), there is a prominent candidate of a group at redshift 0.21. We note that the SDSS data is sharply cut in the lower left corner of Figure \ref{fig:lss}-(a), because it is not observed in SDSS, but fortunately, this does not affect the area of interest around the possible filament feature.

It is interesting to note that the group candidate (Blue rectangular in Figure \ref{fig:lss}-(a)) is roughly on the opposite side to the filament feature. It is well known that filaments form conduits by which groups enter clusters. Therefore, it is possible that this group entered the cluster from the filament and crossed the cluster, emerging on the south side, where it is observed.

\section{Discussion} \label{sec:dis}

In this section, we discuss how our results on the dynamical state of A2261 fit in with previous multi-wavelength studies of A2261, and the fossil origin controversy of whether fossils are transitional or not.

\subsection{Comparison with previous studies about dynamical state of A2261 \label{subsec:prevA2261}} 
In previous researches, there were contradictory results about the dynamical state of A2261. Some consider it as a relaxed system \citep[e.g.][]{2013MNRAS.433.2790L, 2012MNRAS.420.2120M, 2008ApJS..174..117M,2017MNRAS.472.1972H, 2013MNRAS.436..275W, 2016MNRAS.460..569R}. Some others argue that it is disturbed system \citep[e.g.][]{2009MNRAS.392.1509G,2019A&A...622A..24S}. We here summarized studies about weak lensing, strong lensing, radio and X-ray analysis of the dynamical state of A2261. 

The weak and strong lensing studies show the dark matter distribution of A2261, and calculate an accurate mass concentration parameter. Normally, a stable system is expected to have a high mass concentration parameter, and this is also considered as evidence of early formation \citep[see][]{2009ApJ...694.1643U,2002ApJ...568...52W}, because the concentration of dark matter halos evolves with time. By calculating the formation time, \citet{2012ApJ...757...22C} hypothesize that A2261 has not finished forming, and is a borderline relaxed and borderline cool-core system. 

In a recent radio paper, \citet{2019A&A...622A..24S} report a large diffuse radio halo at about 500kpc on the north-west side of the A2261 center. Usually, the radio halo is believed to be formed by major merger, but A2261 is considered as a non-major merger radio halo from the well-ordered X-ray emission contour. If it had a merger before, the X-ray emission contour is expected to be misaligned, because the hot gas of the cluster is disturbed by the infalling matter.

For the X-ray parameters, previous studies divide clusters into different dynamical states based on their X-ray central entropy with \textit{`30$KeV cm^{2}$'} being the dividing line. Lower than 30$KeV cm^{2}$ is categorised as a \textit{`cool-core'} system and larger than 30$KeV cm^{2}$ as a \textit{`non-cool core'} system. However, actively merging clusters have up to 100$KeV cm^{2}$ entropy values. A2261 has moderate X-ray central entropy value, $K_0$(central entropy)=61.08 $KeV cm^2$ \citep{2009ApJS..182...12C}, which is an abnormally high value for a relaxed cluster.

\citet{2017MNRAS.466..996S} argued that they could not find a prominent merger sign. The moderate X-ray entropy feature could indicate a minor off-axis merger. \citet{2016MNRAS.460..569R} hypothesized that A2261 could be at the beginning of a high mass-ratio merger before the first peri-center using their results of low X-ray temperature and SZ and X-ray parameters.

Summing up the previous results, most studies conclude that the formation of A2261 might not yet be complete, and some studies suggest it is undergoing an off-axis minor-merger. An off-axis minor merger could simultaneously explain both the moderately enhanced X-ray entropy value and the well-organized shape of the X-ray contour.  

We propose that the clump in the southern parts of the cluster center, detected by X-ray and radio wavelengths, could be a good candidate for the perturbing galaxy group. We detect this group with our spectroscopic data using a D-S test. Our study also agrees well with previous results on the dynamical state of the cluster. We speculate that one reason for the contradictory dynamical state of A2261 is the off-axis minor merger, which would cause only weak changes in the dynamical state parameters.

Given that the brightest galaxy within the group (m2$\_$all), which is similar mass to the BCG, this further demands for the present state of A2261 is that the merger be off-axis. If an on-axis merger occurred, we might expect it to heavily disturb the BGG by the strong tidal interaction of the two massive galaxies. However, when we check the optical morphology of the BCG and m2$\_$all, we do not find any optical features (such as tidal disturbances or merger features) to support this. Some studies report that A2261 has a flat core structure \citep{2012ApJ...756..159P} and 4 stellar knots inside \citep{2017ApJ...849...59B}, that could be interpreted as boxy isophotes. However, these features were likely made from a much earlier major merger that fully coalesced during the early cluster formation, rather than the recent minor group passage we consider here. Also, If it is passed the center of the cluster, the group has to be broken \citep{2020MNRAS.498.3852B}.

\subsection{What does A2261 tell us about possible transitional state fossils?  \label{subsec:cluetrans}} 

There is much-competing evidence on each side for observed fossil origins, either due to \textit{`real physics'} or as \textit{`by chance (transitional)'} objects. 

First, the \textit{`by chance'} camp argues that fossils and non-fossils have the same BCG parameters \citep{2006MNRAS.369.1211K,2009AJ....137.3942L}, scaling relation \citep{2007IAUS..235..214K}, size-luminosity relation, fundamental plane, Faber-Jackson relation \citep{2012A&A...537A..25M}, absence of large gradient in the metallicity of BCG \citep{2013A&A...553A..99E}, similar amount of substructure \citep{2016A&A...586A..63Z}, and fossils exist in both poor and rich environment \citep{2011MNRAS.417.2927P,2012A&A...540A.105A}. They argue that these same physical properties mean the same formation sequence as in non-fossil systems. 

Especially, the claim of \citet{2013A&A...553A..99E} is critical. A monotonic collapse of the BGG is expected to make a steep metallicity gradient, and only major mergers are able to make flatted metallicity gradient. As a result, a BCG in a fossil cluster should not be able to maintain a high metallicity gradient, if there were multiple mergers of L* galaxies before z=1. However, their six fossil central galaxies showed flat metallicity gradients. In addition, \citet{2014MNRAS.439.2281P} showed two fossil groups with differing star-formation histories, even though both shared similar optical properties and kinematics.

Next, the \textit{`real'} camp asserts that the BCGs of fossil system have more boxy isophotes \citep{2006MNRAS.369.1211K}, fossil system prefer more sparse environments (more closer to filament than node) \citep{2020A&A...639A..97A}, have less bright characteristic magnitude (M*) \citep{2015A&A...581A..16Z}, heavier halo masses than non-fossil at all redshift  \citep{2007MNRAS.382..433D}, lower dwarf galaxy fractions in their luminosity functions \citep{2014A&A...571A..49G}, more radial orbit \citep{2021arXiv210710850Z}, and different assembly bias \citep{2020ApJ...898...39L}. 

Usually, the boxy isophot of BCG is considered as evidence of past major merger. This fits with the L* merger hypothesis to explain the fossil BCG formation. Sparse environments support the passive evolution of the fossil system. Other differences support that a specific process occurred to form the high magnitude gap, and that the above-mentioned features can not be explained as `coincidental results'. 

For the A2261, we search for the above-mentioned features in the literature and catalogs. The BCG of A2261 indeed has a flat core, which can be interpreted as a boxy shape \citep{2017ApJ...849...59B}. Also, it is well-known as a Black Hole-Black Hole merger candidate \citep{2017ApJ...849...59B}, thus, it is usually assumed to have suffered a merger in the past. The X-ray spectral metallicity of A2261's BCG shows a steep decrease within 30 kpc. Beyond that, it becomes flattered until about a 300 kpc radius \citep{2021ApJ...906...48G}. Except for the central part, its metallicity appears to be in quite a mixed state. In the optical luminosity function, A2261 has a less bright M* and a flatter faint-end slope $\alpha$ \citep{2012A&A...540A..90B} (even though, they distinguished A2261 as a disturbed sample, it's M* and $\alpha$ values are more similar with relaxed sample). The sparse environment of A2261 is shown in this paper (see Section \ref{subsec:lss}). Therefore, although A2261's BCG has likely suffered a recent off-axis minor merger, its total structure formation is similar to real fossil formation physics.

Even though we believe that A2261 is likely a transitional fossil cluster, we can not determine that all fossils are transitional using our result. However, we can suppose that a fraction of fossil systems like A2261 might be contaminated samples of pure fossil systems.

\section{Summary} \label{sec:sumncon}

In this study, we attempt to determine the dynamical state of the fossil cluster Abell 2261 (A2261), using our own spectroscopic observations taken with the Hectospec at the MMT to complement existing data in the literature. A2261 is known to have a controversial dynamical state from previous studies of a sample of transitional fossil systems.

We acquired line-of-sight velocities to 373 new galaxies in the field around A2261, and identified 257 member galaxies, and thus improved the spectroscopic completeness within $5R_{vir}$ from the cluster center.

Using the spectroscopic data set, we measured multiple dynamical state indicators such as Gaussianity and luminosity weighted offset indicators, and we found A2261 generally falls on the relaxed side of the distributions of clusters in our comparison sample. 

However, a Dressler-Shectman test (D-S test) confirms with a high significance that substructure exists. One of the probable substructures is located to the south of the cluster, just beyond the virial radius. 
An X-ray emission map highlights that hot gas is associated with this substructure, and has a comet-like shape as if the substructure is interacting with the cluster's hot gas, with a tail pointing towards the cluster. The location of the substructure is well-matched with a bright spectroscopically confirmed member galaxy, m2$\_$all, whose luminosity is within two magnitudes of the cluster BCG. 
A phase-space analysis suggests it is likely located in the cluster outskirts, between one and three virial radii. At its current location, it does not break the fossil criteria but if it had previously crossed the cluster or if it later falls into the cluster, A2261 would transition to being a non-fossil.

Towards the North-East of the cluster, we detect an extended feature of member galaxies out to 5 virial radius from the cluster, that resembles a filament connected to the cluster. Thanks to the addition of our new spectroscopic data, a D-S test reveals a probable substructure within the filament, located at a projected distance of 4$R_{vir}$ from the cluster center. This likely indicates a group is being fed into the cluster along the filament.

Using the available SDSS spectroscopy, we checked the large-scale structure in the surroundings of A2261 until 30$R_{vir}$. We find that the extended feature seen in the Hectospec data becomes a filamentary-like feature of SDSS galaxies on much larger scales ($\sim$15Mpc), that connects with the cluster both in position and in velocity space.

Given that the confirmed substructure to the south is roughly on the opposite side of the cluster from the filament, this raises the possibility that the substructure may have originally been fed into the cluster along the filament, and since crossed the cluster, resulting in an off-axis minor merger, with the group currently being found in the back-splash regions of the cluster today. An off-axis minor merger could explain the reason why many of the dynamical state indicators in this study and previous studies suggest the cluster dynamical state is between relaxed and mildly disturbed. 

Therefore, we suggest that A2261 is in the moderate transitional phase. If the group crossed the cluster, A2261 may have transited from fossil to non-fossil (during the passage of the group across the cluster center on timescales lasting $\sim$1Gyr) and back to fossil (from the moment of transition until now, on $\sim$Gyr timescales). In the future, it will likely return to being non-fossil, when the group falls back into the cluster, on similar timescales. This future transition will likely occur even if the group is in fact on first infall. 

While the study of one system cannot drive broad conclusions on fossil systems as a whole, we suggest that the conventional fossil definition likely includes other transitional systems like A2261 that pollute samples of true, pure, highly relaxed fossil systems.

Therefore, we suggest the need for a moderate classification to represent transitional phase fossils, rather than a dichotomy classification. Different dynamical state indicators likely become perturbed at different stages of an ongoing merger, and remain perturbed for differing amounts of time. We will study this using numerical simulation in the near future, and use the results to give increased knowledge on the recent merger history of observed clusters, and to better identify fossil systems that could be in transition.

\begin{acknowledgments}
We thank the anonymous referee for constructive comments and detailed suggestions that make this paper clearer. We thank Cristiano Sabiu for helpful discussions. We thank Ana Laura Serra and Antonaldo Diaferio for allowing us to use the Caustic App. This project was supported by K-GMT Science Program of Korea Astronomy and Space Science Institute. Hectospec observations used in this paper were obtained at the MMT Observatory, a joint facility of the Smithsonian Institution and the University of Arizona. H.S.H. acknowledges the support by the National Research Foundation of Korea (NRF) grant funded by the Korea government (MSIT) (No. 2021R1A2C1094577). JWK acknowledges support from the National Research Foundation of Korea (NRF), grant No. NRF-2019R1C1C1002796, funded by the Korean government (MSIT).
\end{acknowledgments}


\vspace{5mm}
%\facilities{MMT(Hectospec)}


%\software{Caustic App v1.6 \citep{2013ApJ...768..116S},
%          IRAF RVSAO \citep{1998PASP..110..934K},
%          Specpro \citep{2011PASP..123..638M}
%          }




\bibliography{main}{}

\bibliographystyle{aasjournal}



\end{document}